# Using Computational Essays to Redistribute Epistemic Agency in Undergraduate Science

Tor Ole B. Odden[1], Devin Silvia[2], and Anders Malthe-Sørenssen[1]

[1] Center for Computing in Science Education, University of Oslo, Norway

[2] Department of Computational Mathematics, Science, and Engineering, Michigan State University, USA

## Abstract

This article reports on a study investigating how computational essays can be used to redistribute epistemic agency—cognitive control and responsibility over one's own learning—to students in higher education STEM. Computational essays are a genre of scientific writing that combine live, executable computer code with narrative text to present a computational model or analysis. The study took place across two contrasting university contexts: an interdisciplinary data science and modeling course at Michigan State University, USA, and a third-semester physics course at the University of Oslo, Norway. Over the course of a semester, computational essays were simultaneously and independently used in both courses, and comparable datasets of student artifacts and retrospective interviews were collected from both student populations. These data were analyzed using a framework which operationalized the construct of epistemic agency across the dimensions of programming, inquiry, data analysis and modeling, and communication. Based on this analysis, we argue that computational essays can be a useful tool in redistributing epistemic agency to students within higher education science due to their combination of adaptability and disciplinary authenticity. However, we also argue that educational contexts, scaffolding, expectations, and student backgrounds can constrain and influence the ways in which students choose to take up epistemic agency.

## Introduction

For centuries, science learning was essentially a process of collecting and curating sets of facts (Livingstone, 2010; Rudolph, 2005), and traces of this philosophy still live on in conservative modes of science teaching today (Stroupe, 2014). With the rise of ambitious science teaching, however, this paradigm has shifted towards an educational philosophy that gives students greater control, responsibility, and agency over their own science learning (Miller et al., 2018; Stroupe, 2014; Windschitl et al., 2012). This trend has inspired a wave of science education research on how teachers can empower students to take up this epistemic agency, especially within pre-college science education. For example, the field has seen the development of culturally-relevant and critical pedagogy (Barton & Tan, 2010; Miller et al., 2018), communities of students explaining, debating, and critiquing ideas (Berland, 2011; Berland et al., 2015; Berland & Crucet, 2015; Berland & Reiser, 2011), and, within higher education, a focus on authenticity in scientific laboratory work (Brownell et al., 2012; Cooper, 1994; Holmes, 2020; Holmes & Wieman, 2016, 2018; Kloser et al., 2011). This shift has also been instantiated into curriculum and education standards reform, such as the Next Generation Science Standards (Miller et al., 2018; NGSS Lead States, 2013).

Despite this progress, much of the teaching in higher education STEM remains fixed within a traditional model of instruction that greatly limits the agency students have over their own learning (Stains et al., 2018). However, in recent years a new trend has emerged that may help ameliorate this issue: the increasing adoption of scientific computation (Caballero & Merner, 2018; Young et al., 2019). By *scientific computation*, we mean the use of computational tools (usually computer code) to create and run computational models and data analyses.

We are interested in exploring the ways these computational tools may be leveraged to promote students' epistemic agency. More specifically, we are interested in understanding the opportunities computation may offer for redistributing epistemic agency to students; the tools, techniques, and best practices can help in this endeavor; and the challenges that emerge along the way.

In this paper, we report on the development of one such tool, called a computational essay, and its implementation in two contrasting university contexts: an interdisciplinary computational modeling and data science course at Michigan State University, USA, and a 3rd-semester physics course at the University of Oslo, Norway. Over the course of a semester, computational essays were simultaneously and independently used in both of these universities and courses, and comparable datasets were collected from each implementation. Using this data, as well as our own experiences as practicing computational scientists and educational designers, we have operationalized the construct of epistemic agency for computational science education and have used it to analyze the affordances computational essays provide for redistributing epistemic agency.

## Theory and Motivation

This study builds off the theoretical construct of epistemic agency, which has garnered increasing interest in the science education research literature over the last three decades. Here, *epistemic agency* refers to the ways in which students, teachers, or social groups take on and distribute cognitive authority—that is, the responsibility for creating and evaluating new knowledge (Barton & Tan, 2010; Damşa et al., 2010; Scardamalia, 2002; Scardamalia & Bereiter, 1991; Stroupe, 2014). When students take up epistemic agency, they take control and ownership of their own processes of learning and inquiry, becoming *epistemic agents*: "individuals or groups who take, or are granted, responsibility for shaping the knowledge and practice of a community" (Stroupe, 2014, p. 492).

Our conceptualization of epistemic agency is built on three foundational pillars of educational philosophy taken from research in science education and the learning sciences.

1. **Epistemic agency is tied to authentic scientific practices**

Drawing on the work of Ko and Krist (2019), Stroupe (2014), and Lehrer and Schauble (2007), our conceptualization of epistemic agency is built on a practice perspective of science learning. This perspective posits that learning science involves learning the key theories and principles that scientists use to make sense of the world, the epistemological foundations that they use to evaluate knowledge, social practices for using and communicating ideas, and the tools and resources needed to do scientific work (Stroupe, 2014). Thus, one key element in supporting students' epistemic agency is enculturating them in practices used by actual scientists, such as experimental designs (Brownell et al., 2012; Kloser et al., 2011), aids to measurement (Dounas-frazer et al., 2016; Dounas-Frazer & Lewandowski, 2016), mathematical and statistical models (Watkins et al., 2012), laboratory practices (Cooper, 1994; Dounas-Frazer & Lewandowski, 2018; Holmes, 2020; Sandi-Urena et al., 2012), scientific communication (Blakeslee, 1997; Kloser et al., 2011; Moskovitz & Kellogg, 2005), and scientific computation (Magana et al., 2016; Odden et al., 2019). Ideally, these practices will be acquired within a social community wherein students will be able to share results, questions, difficulties, and critiques with one another, analogous to the professional communities formed by practicing scientists (Irving et al., 2017, 2020; Ko & Krist, 2019; Lehrer & Schauble, 2007; Stroupe, 2014).

2. **Epistemic agency requires opportunities for scientific inquiry**

Our second pillar relates epistemic agency to the construct of scientific inquiry. Building on the work of Chinn and Malhotra (2002) and Ko and Krist (2019), we assert that real agency—that is, having some degree of actual control of learning rather than the illusion of control—requires some degree of open-endedness in the problem under investigation (Chinn & Malhotra, 2002; Holmes, 2020; Wieman, 2015). To be epistemic agents, students need the freedom to choose the direction of their project or investigation and change that direction as new information is discovered or challenges arise—making decisions for how to proceed, how to navigate roadblocks, and when to stop or continue (Holmes et al., 2020; Ko & Krist, 2019). They must also have the freedom to solve challenges in different ways depending on preference, knowledge, and skills. Thus, another key element in supporting students' epistemic agency is positioning them such that they have opportunities to engage in scientific inquiry, especially on questions that they find interesting and relevant (Chinn & Malhotra, 2002).

3. **Epistemic agency is tied to shared knowledge creation**.

Building on the work of Damşa et al. (2010), we see epistemic agency as theoretically tied to a knowledge creation perspective, the idea that "learning takes place through collaborative activities whose aim is to create new knowledge through work on shared objects" (Damşa et al., 2010, p. 146). Thus, epistemic agency involves students being active participants in a social knowledge-construction process, within which the participants have cognitive responsibility for managing and monitoring their own learning (Scardamalia, 2002; Stroupe, 2014). When taken to its limit, this perspective advocates that students take responsibility not just for their own learning, but also the decision of what to learn, how they are going to learn it, and how they will judge if they have been successful or not (Stroupe et al., 2018).

**Epistemic agency and scientific computation in higher education**

Given this relationship between epistemic agency and authentic scientific practice, one might expect agency-based teaching to be the default in higher education STEM. After all, students in higher education STEM are being trained as professionals in a specific scientific discipline; they are usually learning from practicing members of that discipline; they have the opportunity to engage in advanced scientific work using the tools of the discipline; and many have opportunities to engage in undergraduate research. However, in practice, all three of the above-mentioned pillars of epistemic agency are often suppressed or absent from higher education science, as can be seen from the continued efforts at large-scale curricular reform across STEM disciplines (Cooper & Klymkowsky, 2013; Henderson et al., 2011, 2012; Thompson et al., 2013) and various national reports (American Association For The Advancement of Science, 2011; Association of American Universities, 2017; Olson & Riordan, 2012). In part, this is a function of the fact that traditional teaching methods, which emphasize rote, closed-ended problems and lecture-based instruction are still widespread at the postsecondary level (Stains et al., 2018). There is also the philosophical challenge that as one advances to higher levels within a scientific discipline, the questions one can ask (and phenomena one can investigate) become increasingly constrained by the theory, tools, and previous work done in that discipline. For example, in more advanced physics courses it quickly becomes apparent that only a small number of problems can be solved using traditional, analytical methods.

One potential solution to this problem, which has emerged in recent years, is the increasing adoption and use of computation for teaching and learning science. This adoption has developed concurrently with the advent and spread of the various sub-disciplines of computational science. *Computational science* (not to be confused with the discipline of *computer science*, or as it is sometimes known, informatics) is the use of computation within scientific fields to solve scientific problems as exemplified by the disciplines of bioinformatics, computational physics, computational chemistry, and data science (Denning & Tedre, 2019).

Computation offers numerous benefits for higher-education STEM. Computation is flexible: it can be used across many different disciplines and problem domains. Computation is cheap: most widely-used programming languages are available for free and are frequently updated with new tools, methods, and packages. Computation is scalable: computational simulations can be safely built and run on most computers, or even mobile devices with the advent of cloud computing. Open-source software packages also make more complex tools and datasets used by professionals available to students at little or no cost. Many different types of analyses can be done with a relatively small set of fundamental techniques such as numerical integration and regression. Once acquired, these techniques can then be used in many different applications across a wide range of scientific disciplines. Thus, with the right preparation, students now have unprecedented access to some of the most cutting edge tools in modern science.

We are interested in exploring the ways computation can be leveraged to create more opportunities for epistemic agency in higher education STEM. As both educators and practicing scientists ourselves, we see great opportunity in using computation to help our students gain authentic scientific skills, engage in genuine cycles of scientific inquiry, and create a knowledge-building community.

To explore these possibilities, we have implemented and tested a teaching tool called a computational essay (diSessa, 2000; Odden et al., 2019; Wolfram, 2017) within two separate university contexts, collecting comparable data from both. Below, we describe the key features of computational essays which make them useful tools for supporting student agency and the contexts within which the study took place.

## Educational Design and Context

### Computational essays: An emerging genre

Computational essays are a new genre of writing that has emerged within the last two decades, whose key innovation is the blending of narrative text with live, executable computer code in order to present an argument, explain an idea, describe an analysis, or tell a story (diSessa, 2000; Odden et al., 2019; Odden & Burk, 2020; Odden & Malthe-Sørenssen, 2021; Wolfram, 2017). A typical computational essay shares many of the features of a traditional essay: a title, introduction, thesis statement, body, discussion, conclusion, references, and possibly graphics, images, or illustrations. However, in a computational essay, a key piece of the essay's argument comes in the form of blocks of code which simulate a phenomenon or analyze a dataset. The outputs of these code blocks are woven together with explanatory text to form the overall narrative of the essay. An excerpt from a student-written computational essay is shown in Figure 1.

**Figure 1:** Example of a computational essay from MSU.

Computational essays are frequently used by practicing scientists, data analysts, and technical professionals to share results and analyses or illustrate computational techniques (Kluyver et al., 2016; Rule et al., 2018; Somers, 2018). They are often written in computational *notebooks*, programming environments that allow users to combine blocks of executable code with text, equations, images and embedded videos. One of the most popular types of notebooks is the open-source Jupyter notebook (Kluyver et al., 2016), which supports over 40 programming languages and was the programming environment used for all computational essays included in this study. Other tools that support computational essays include Mathematica (Wolfram, 2017) and the MATLAB Live Editor, both of which are available under commercial license.

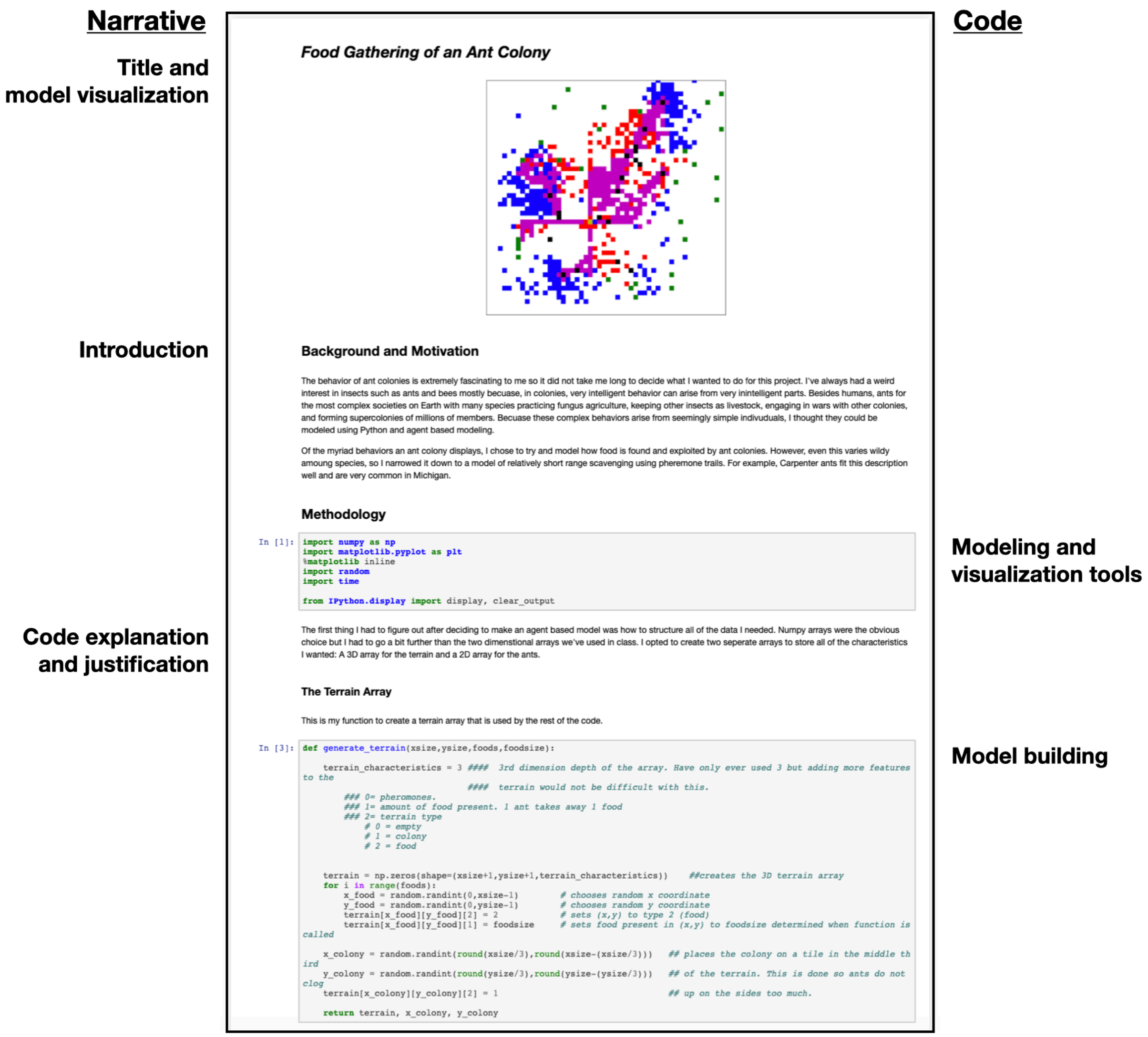

Computational essays, by their nature, provide numerous affordances for supporting students' epistemic agency as described in the three pillars above. First, the process of writing a computational essay is an authentic scientific practice. In order to write a computational essay on a scientific topic or problem, one must use code to perform a computational analysis or simulation (a key scientific practice) and explain the steps, results and relevance of that simulation in writing (another practice). Furthermore, the goal of a computational essay is to convince a reader of the validity of some argument or analysis, which provides opportunities to develop the practice of scientific argumentation (Berland, 2011; Berland & Reiser, 2011).

Second, computational essays can help scaffold open-ended, inquiry-based work (Odden & Malthe-Sørenssen, 2021). Students can write computational essays on a myriad of topics, potentially based on their own interests or completely novel questions. They also can choose how they wish to answer the question, how to address challenges or roadblocks along the way, and how much evidence to provide in support of their argument. In this way, computational essays incorporate certain elements of writing assignments from the humanities, where students have the freedom to construct an argument and the evidence to support it, while also retaining the replicability of computational science simulations.

Third, computational essays are a genre designed for communication and collaboration, providing an ideal medium for shared knowledge creation. Computational essays allow students to share their ideas, analyses, and critiques in much the same way that professional scientists do. Because students can collaborate on both computational analyses and writing, they thus provide an ideal "shared object" for the building of new knowledge (Damşa et al., 2010). The portability and modifiability of code also opens the potential for students to test, reuse, and build off of each other's work, a key element of a knowledge-building community.

Despite these many potential benefits, there has as yet been little study on the use of computational essays in education. There has also been little attention to the possibilities of using either computational essays or computation more generally for supporting student epistemic agency in STEM education. Thus, our research questions are as follows:

1. **In what ways can computational essays support the redistribution of epistemic agency to science learners?**
2. **How do different implementations of computational essays affect the ways in which learners take up epistemic agency?**

**Computational essay implementation**

Based on these research questions, we studied the use of computational essays in two separate university contexts: an interdisciplinary computational modeling and data science course from Michigan State University, USA (hereafter MSU), and a 3rd-semester physics course from the University of Oslo, Norway (hereafter UiO). Both of these courses independently used computational essays during the fall semester of 2019. Although the authors of this paper, who are also instructors and designers for the respective courses, had been in contact prior to this semester, at the time of data collection the educational designs (which had been developed during previous semesters) were already established. That is to say, no significant changes were made to these educational designs for the purposes of this study.

At the University of Oslo, computational essays were used as the basis for a mid-semester computational modeling project, situated roughly two-thirds of the way through the semester (Odden & Malthe-Sørenssen, 2021). The goal of the project was to give students the opportunity to try their hand at defining and investigating a problem using computational modeling, then present their results to their peers in a scientifically-authentic way (computational essays, followed by oral presentations). Prior to this course, most participating students had taken an introductory course in object-oriented scientific programming, as well as one or more physics and math courses that included a computational modeling component. However, this project was the first time most students had had the freedom to define their own investigation question. Projects were done singly or in pairs, and were ultimately graded pass/fail using a lenient grading system meant to incentivize student agency.

At Michigan State University, computational essays were used as the basis for an individual, end-of-semester, project-based summative assessment for an introductory, interdisciplinary data science and modeling course. Students in this course were not expected to have any previous programming experience, and over the semester they had learned basic Python programming skills; how to build, run, and interpret computational models; and how to manipulate, analyze, and visualize datasets (Silvia et al., 2019). Notably, every assignment in the course was written in a Jupyter notebook, and the course included explicit and implicit instruction on code documentation and notebook use. All projects included a presentation component wherein students orally presented their results to their peers using their essays as the basis for their presentation slides. Projects were graded using a rubric, also provided to the students, that emphasized computational modeling and communication of questions and results.

The two implementations shared certain key features. Both required students to define a question that would try to answer using a computational model or analysis; create a model that could be applied to their dataset or chosen topic; and use computational methods to answer their question. Both required students to communicate their question and analysis in the computational essay format, and use that computational essay as the basis for an oral presentation to their peers. Both implementations also provided students with several forms of scaffolding in the computational essay-writing process, such as an assignment description, a copy of the rubric used to grade the essays, example essays written by staff or previous students, and help-sessions run by instructional staff. And, both implementations placed significant weight on the processes of inquiry, computational modeling, and scientific communication, rather than canonical correctness of models (which was not included in either assessment rubric).

However, there were also certain key differences in implementation. For example, at UiO all projects were physics-focused, since computational essays were situated within a physics course, and students were explicitly required to use physics principles from the course within their projects. Students at UiO had the option to do the project in pairs, whereas all MSU students completed their projects individually. As compared with MSU, at UiO students were provided significantly more example essays, written by faculty and previous students and posted in a publicly-accessible online showroom (Center for Computing in Science Education, 2019). Whereas MSU students were all required to define novel investigation topics, at UiO students were provided with suggestions for questions to pursue in case they had difficulty coming up with a question on their own. These questions were accompanied with “seed” programs that they could build off if they chose, which consisted of stripped-down simulations of physics phenomena that were specifically designed to run without error but provide limited insight into the physical system in order to give students opportunities for exploration.

At MSU, student backgrounds varied considerably, and so essay topics covered subjects from the natural sciences, social sciences, statistics and mathematics, and beyond. MSU students were provided with fewer example essays but were given a template notebook to build off of which specified the structure of the resulting essay (including headers like Background and Motivation, Methodology, Discussion and Conclusions, and References). Students at MSU were also given more oversight throughout their projects: prior to starting their project, MSU students were required to submit a project proposal describing the question they planned to investigate and were given feedback on the feasibility of their chosen topic. A few weeks before the project deadline, students then presented a project update to their peers showcasing what they had accomplished up to that point. Project updates not only served as a motivator to get students to begin work on their projects, but also highlighted to instructors if students were on a path that might not lead to positive project outcomes. Essays from MSU also tended to come in two “flavors” or sub-genres: *modeling essays*, which used a computational or mathematical model to investigate a physical phenomenon, and *data-driven* *essays* which used statistical and computational techniques to analyze large datasets, whereas essays from UiO were exclusively in the modeling sub-genre.

Additional details on the course contexts and computational essay implementations can be found in (Silvia et al., 2019) and (Odden & Malthe-Sørenssen, 2021).

## Data Collection and Analysis

Because of the wide variation in student backgrounds and interests in the study population, the international nature of the study, and the novelty of computational essays within an educational context, this study used a mixed-methods approach. Our goal with this approach was to both capture broad trends in how students approached the computational essay project within the two surveyed contexts and understand student thoughts, feelings, and experiences when doing this kind of open-ended, project-based computational work. Accordingly, we gathered the following forms of data:

**Artifacts:** Completed computational essays from all consenting students or pairs of students. 45 of these essays came from MSU, and 58 from UiO. All MSU essays were in English and done by individual students. At UiO, 49 of the 58 essays were in Norwegian, and 20 of the 58 were from students working in pairs. All essays were anonymized by deleting any names or identifying information. The majority of artifacts collected from MSU students were data-driven essays, while UiO essays were exclusively model-based.

**Interviews:** In addition to collecting completed essays, we interviewed a subset of students (14 from MSU, 12 from UiO) using a similar interview protocol at both institutions to try to capture students' motivations and inquiry processes in their own words. All interviews were retrospective, held around the time students completed and presented their projects. The protocol included questions about the students' academic and programming background; their comfort with programming; their reflections on the ways programming could contribute to science learning; a description of the way they approached the computational essay writing task as well as their motivation for their project topic and challenges they faced; and their explicit reflections on feelings of creative freedom and ownership over the learning process. All interviews also included a component in which students walked the interviewer through the different parts of their essay (displayed on the students' laptop). During recruitment, we made an effort to capture a wide variety of student backgrounds, demographics, and project topics. All MSU interviews, and two of the UiO interviews, were in English; the remaining 10 interviews were in Norwegian. All interviews were transcribed and anonymized (using pseudonyms), although interviewed students were linked to their specific essays in the artifact dataset.

**Data Analysis**

Data analysis proceeded in several stages. We began by creating an analytical framework for epistemic agency, in the form of a codebook that could be applied to the collected computational essays. We based this codebook on the epistemic agency literature, our own design philosophies/decisions in the courses, and our experiences as professional researchers in computational science and data science. Our explicit goal with this codebook was that it would capture most of the ways in which students *could* display epistemic agency in their finalized computational essays. That is, we were aiming to "cast a wide net" in order to capture interesting distinctions and nuances in the ways students approached the task, knowing that different students would prioritize different elements (such as programming, investigation question, modeling, and writing or presentation), and also realizing that the two institutions likely supported students in different ways. For this reason, we designed the codebook on a scale of 0–2, with 0 being deficiency in that category, 1 being sufficient to receive a passing grade according to the respective rubric, and 2 indicating that the student had gone above-and-beyond the course expectations and taken up epistemic agency within this category. Our final codebook included the following categories (elaborated in Table 1):

**Programming and Data Processing (P):** This category focused on ways in which students took up agency through the practice of scientific programming. Specifically, some students spent effort on making their code run efficiently and elegantly (subcategory P1), which is an important practice when writing long or complex computational simulations. Other students made a visible effort to make their code organized, readable, and documented (subcategory P2). Still other students took ownership over the coding process by using programming tools and packages in their projects that had not been covered within their courses (subcategory P3).

**Investigation (I):** This category focused on ways in which students took up agency through the scientific inquiry process. Specifically, some students spent effort on defining novel questions for their investigations (subcategory I1), which is a critical part of scientific inquiry.

Other students made an effort to develop their investigations beyond initial, surface-level results, resulting in multi-step investigative narratives (subcategory I2). Still other students made a clear effort to provide detailed discussion and interpretation of the results of their analyses, including reflections on understandings gained or connections with real-world systems (subcategory I3).

**Modeling and Data Analysis (MD):** This category focused on ways in which students took up agency through the processes of scientific modeling and data analysis. Specifically, some students paid explicit attention to the sophistication and realism of their model or analysis (subcategory MD1). These students often produced unique visualizations and/or detailed analyses of their chosen phenomena. Other students paid explicit attention to addressing limitations or assumptions inherent to their modeling or data analysis techniques (subcategory MD2), a key part of any scientific endeavor. Still other students spent effort consulting multiple external sources and using these to improve their models or analyses (subcategory MD3).

**Communication (C):** This category focused on ways in which students took up agency through the scientific communication process by fully using the communicative capabilities of computational notebooks. Specifically, some students spent extra effort in explaining and justifying the meaning and structure of their code in the text of their reports (subcategory C1). Other students spent visible effort in polishing their written reports, sometimes deliberately writing in a particular genre (like scientific paper or personal essay) (subcategory C2). Still other students made a deliberate effort to use graphics, plots, and illustrations in their written reports to either communicate key findings or simply generate visual interest (subcategory C3).

Taken together, these categories and sub-categories operationalize the three pillars of epistemic agency discussed above within the context of scientific computation. In reference to the practice pillar, all four categories represent authentic scientific practices that are used whenever professional scientists perform and present computational analyses. In reference to the inquiry pillar, inquiry was deliberately included as a category within the rubric, and we paid explicit attention to the ways in which students defined and pursued their investigation questions. In reference to the final pillar of shared knowledge creation, the definition, investigation, and presentation of a novel question to a reader represent a clear example of this kind of social knowledge construction. Finally, across all these categories, the primary focus was on visible markers that students had made a deliberate effort to take ownership over this category, rather than judgments on correctness of contents or practices.

| Code Category | Code Abbreviation | *Description* | Level 0 (no evidence) | Level 1 (limited/basic) | Level 2 (advanced) |
|---|---|---|---|---|---|
| **Programming and Data Processing** | **P** | *Students taking ownership of coding and data processing by going above and beyond the default expectations. Focus here is on the blocks of code and their organization.* | | | |
| **Code efficiency and elegance** | **P1** | *Use of code structures and compartmentalization methods such as functions and classes to make code more elegant and efficient* | Little or no code. Code very inefficient or inelegant. | Only basic/default methods used: variables, loops, calls to packages, pre-defined functions. Lots of copying and pasting. | Effective and efficient use of functions and classes. Code organized efficiently and elegantly to avoid excess copy/pasting or re-definition. |
| **Code organization and self-description** | **P2** | *Use of notebook cells as an organizational tool to make code more readable, and use of variable names and comments/docstrings to make code understandable.* | Code is all lumped together into one chunk and hard to read. Little-to-no use of comments or docstrings. | Code is split somewhat across the notebook, but still heavily lumped together. Limited-to-moderate use of comments and docstrings. | Code is logically split into digestible chunks, sequentially organized. Comments and/or docstrings are used appropriately and effectively. Variable and function names are logical and communicate the idea behind the code. |
| **New packages and coding tools** | **P3** | *Students using new packages and coding tools that they had to learn on their own, appropriate to the disciplinary focus* | No packages or coding tools used. Bare minimum required to run code (or no code present at all) | Default packages and coding tools used (those featured in the course) | Any new packages and coding tools (including modeling or visualization tools) used, and/or students define/create a new computational package for their project |
| **Inquiry** | **I** | *Students taking ownership over the scientific inquiry process by defining novel research questions, developing investigations in novels ways, and making novel interpretations* | | | |
| **Investigation question** | **I1** | *Students define a novel research question* | Research question unclear | Standard/suggested research question and/or research question is overly | Research question is well-defined and clear in scope. If it is a novel, unusual, or personally relevant research |
| | | | | broad, not well-defined. | question, this makes up for lack of definition/clarity. |
| **Investigative narrative** | **I2** | *Students developing their models and/or analyses beyond their initial results* | No investigative narrative; all analysis done in a single step, or investigation is inserted into an otherwise-unchanged set of pre-written code. | Investigative narrative is present, but stops at initial results | Multi-step investigation, including refinements on initial results |
| **Discussion and interpretation of results** | **I3** | *Students making novel interpretations and/or describing new ideas they discovered along the way* | No interpretations | Limited interpretations, students just restate results. | Students provide a summary of the results of the project. Detailed discussion and interpretation of results. Possible reflection on new understandings gained, connections to the real world, or alignment with initial expectations |
| **Modeling and Data Analysis** | **MD** | *Students taking ownership of the scientific modeling and data analysis process* | | | |
| **Model development and data exploration** | **MD1** | *Students taking control of the model construction process and/or the exploration of their data* | Students just modify parameters of a pre-existing model, make the very basic plots of the most easily accessible variables in their data, and/or consider most surface-level variables. | Students construct a basic new model with significant resemblance to things they've already done or learned about, and/or make minimal modifications to an existing model. Analysis goes beyond surface level variables, but includes limited data exploration. Some use of more advanced data visualizations to extract | Students build a sophisticated new model from the ground up and/or extensively modify or adapt an existing model. Students visualize the data in unique ways that provide critical insight into their understanding of the data (may include types of plots not covered in the course content). Students do a comprehensive exploration of data, including several different kinds of comparisons, and/or examining controlling variables |
| | | | | additional information. | within the given dataset. |
| **Assumptions and limitations** | **MD2** | *Students considering the limitations of their models or their datasets and addressing them* | Limitations not considered or addressed | Assumptions and/or limitations only superficially considered (just noted/stated), little attention to addressing them or unpacking their implications for the results. | Limitations carefully and thoroughly considered and addressed in the model. Students discuss the assumptions inherent to their model or analysis technique, limitations in their modeling technique or dataset, and/or how they might improve the model or analysis by addressing assumptions and limitations. |
| **External sources** | **MD3** | *Students consulting external sources for inspiration and evaluation of model* | No external sources consulted or cited | Small number of sources consulted (1-3), and/or mostly general-purpose sources (i.e., wikipedia) | Large number of sources consulted (> 3), and/or sources taken from specific domains or literatures (i.e., reports, papers, or articles) |
| **Communication** | **C** | *Students taking ownership over the scientific communication and presentation process, and fully using the communicative capabilities of the notebook for this purpose* | | | |
| **Code explanation and justification** | **C1** | *Students going above-and-beyond in the narrative text explanations of their code* | Little or no writing, just code | Some explanation of code throughout the notebook, but limited to simple descriptions of what the code does and/or basic comments | Detailed prose, fleshed out and persuasively written throughout the notebook. Students not only explain what code does, but also how it works and why it is necessary (motivates the code). |
| **Writing genre and polish** | **C2** | *Students explicitly trying to write their essays in a particular genre, such as a scientific report or personal essay, and attending to the quality of writing in* | Report is very basic, poorly written, and/or lacking. Little or no writing, and/or prose is limited to very basic explanations of what the students | Report is adequately written. Writing is mostly limited to explanations of what they did, but students have introduced some structure and organization into their report | Report is well-written and polished. Clear markers that the students have tried to write their essays for in a particular genre/style, such as use of informative section headers for organization |
| | | *their essay* | did | | |
| **Graphics, images, and illustrations** | **C3** | *Use of graphics, images, and animations to illustrate key pieces of the essay and/or set the tone for the work* | Limited-to-no use of graphics, illustrations, animations, or graphs and plots beyond the most basic graphs needed to communicate results. Students have not produced anything new. | Moderate use of illustrations; visualizations, graphics, and plots; enough to communicate both the project design and the project results, with visualizations sprinkled throughout the report. Limited consideration to aesthetics or understandability. | Students have put significant effort into the visual design of the computational essays. Frequent use of illustrations, graphics, animations and plots to communicate ideas, key points, and/or improve aesthetics of essay. Clear thought put into the colors, organization, or other elements to make it more aesthetically appealing and/or understandable. |

**Table 1:** Epistemic agency codebook developed for analysis of computational essays

Once we had created an initial draft of this codebook, our analysis proceeded as follows: First, we engaged in iterative coding on a subset of the data, composed of roughly 20% of the essays (19 in total: the 9 English essays from UiO plus 10 randomly chosen essays from MSU) in order to refine our codebook and establish initial estimates of inter-rater reliability. These initial rounds of coding resulted in several small-scale changes and clarifications to the codebook, and also revealed the necessity of sharing additional information about the course contexts between coders to clarify some of the evaluations made.

To evaluate inter-rater reliability, we initially used the standard Cohen's Kappa statistic. However, after inspection of the results from the initial rounds of coding, we realized that despite substantial percent agreement (>70%) across hundreds of ratings in different categories, the IRR statistic was low (< 0.6). Upon further investigation, it turned out that the value was being suppressed by the "paradox" inherent to Cohen's Kappa, in which categories with high agreement will produce low Kappa values (Feinstein & Cicchetti, 1990). For example, categories that produced nearly 100% agreement caused the Kappa statistic to go to 0. For this reason, we instead chose to use a "paradox-resistant" IRR statistic, Gwet's AC2 (Gwet, 2014), benchmarking the degree of inter-rater reliability using the procedure described by Gwet (2014).

Using AC2 as our metric, the first two authors, TOBO and DWS, then repeatedly coded and modified the codebook until we reached "very good" agreement based on this benchmarking scale (AC2 > 0.85). Thereafter, the dataset was split based on institution and every remaining essay was independently coded by two researchers, with discrepancies being resolved through discussion. All remaining MSU essays were independently coded in batches by DWS and TOBO. Prior to discussion, percent agreement ranged from 72% to 86% (increasing with each batch) and AC2 scores ranged from 0.87-0.9, consistently meeting the “very good” threshold of agreement. All remaining UiO essays were independently coded in batches by TOBO and a graduate researcher who had previously taught other computationally-focused courses in the UiO physics department. Prior to discussion, percent agreement ranged from 67% to 75%, with AC2 scores ranging from 0.74-0.88, reaching the “good” and “very good” thresholds of agreement.

After all essays had been analyzed, we performed several secondary analyses on the resulting codes. These included examinations of overall levels of agency across multiple dimensions: aggregate, split by institution, and split by type of essay (modeling vs. data analysis). In some cases, the trends revealed by these analyses seemed connected to the specific design decisions made in the computational essay implementations described above. Additionally, we performed an association analysis on our codes, both aggregated and split by institution, using the bias-corrected Cramér’s V statistic (Bergsma, 2013; Cramer, 1999; Kearney, 2017), and looked for ways in which these associations could be connected either to overlaps in the codebook or course design. Finally, we used a subset of the codes to create narrative case-studies of specific interviewed students, in order to both validate our codebook with interview data and illustrate different ways in which students demonstrated agency through this project. To do so, we isolated the codes for all interviewed students and grouped them into bins of low, medium, and high agency. We then chose representative students from each bin (aiming to present a mix of narratives across institution, gender, and group composition) and reviewed their interviews and essays to try to reconstruct the story of their essay-writing process and capture their reflections on creativity and agency within the project.

## Results Part 1: Case studies of varying levels of epistemic agency

We begin with the narratives of three groups of students who showed differing levels of epistemic agency according to our analysis: low (total score 11-12), medium (14-16), and high (21 or above). These cases are meant to both provide both an illustration of key features of computational essays and showcase the different ways students took up agency through the assignment. For context, Figure 2 shows a heat map of all computational essay scores, disaggregated by institution and color-coded according to level of demonstrated epistemic agency (darker = higher score). In this figure, we have highlighted the three groups whose narratives are featured here. We also summarize key features of these three cases in Table 2.

| Student Pseudonym(s) | Student University and Major(s) | Epistemic Agency Score | Computational Essay Topic | Categories of Demonstrated Agency |
|---|---|---|---|---|
| Amy and Alexis | UiO; Electronics, Informatics and Technology | 11 (Low) | Lightning safety in cars | External sources; Writing genre and polish; Graphics, Images, and Illustrations |
| Iggy | MSU; Computer Science | 14 (Medium) | Baseball statistics | Investigation Question; External sources |
| Margaret and Edward | UiO; Materials Science and Physics | 22 (High) | Railgun dynamics | All categories **except** Code efficiency and elegance, and New packages and coding tools |

**Table 2:** Summary of case studies of low, medium, and high agency groups

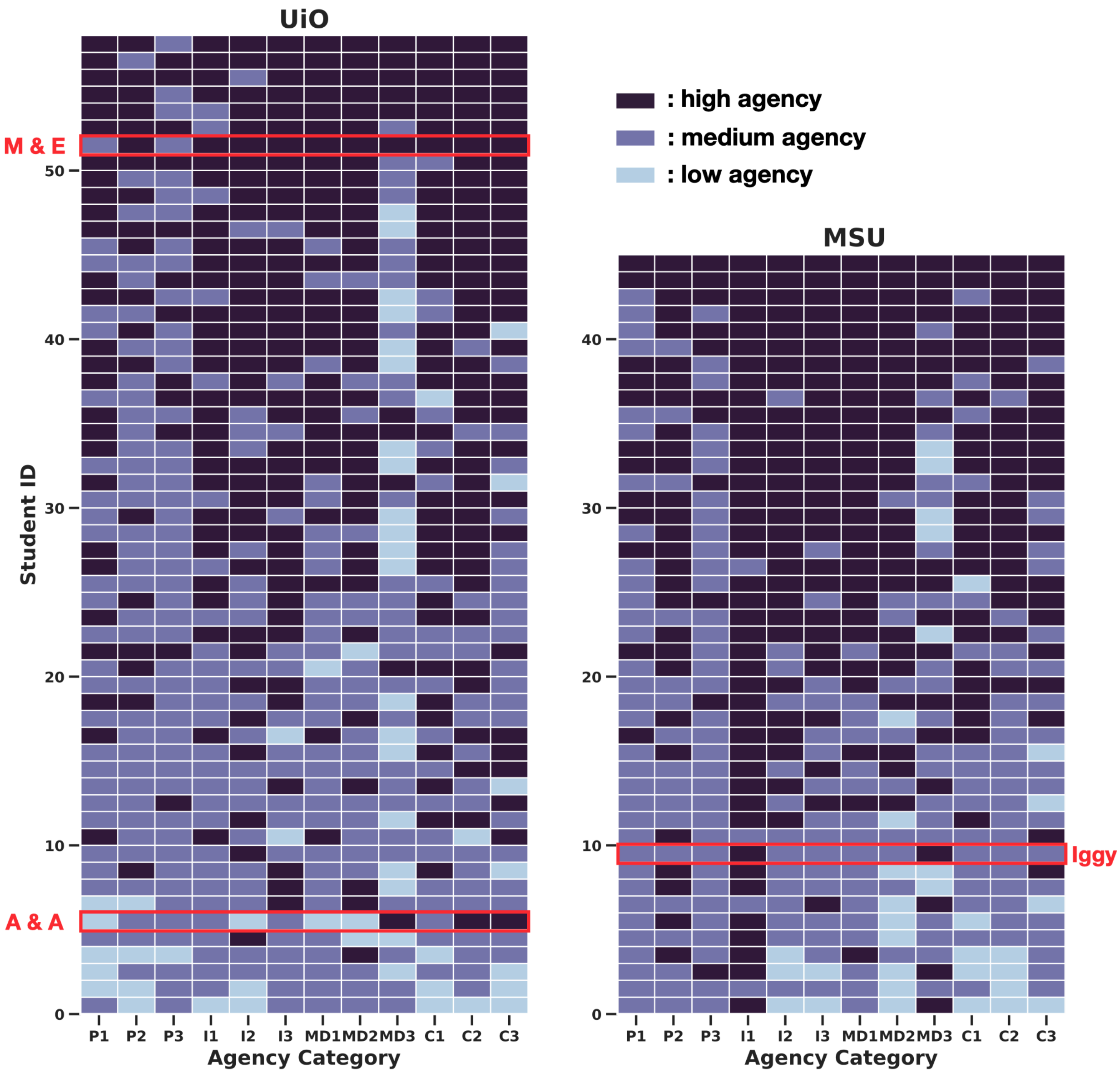

**Figure 2:** Heat map of epistemic agency codes, disaggregated by institution and highlighting students featured in narratives. Darker colors indicate higher levels of demonstrated agency.

**Low agency: Amy and Alexis focus on communication at the expense of computation**

Amy and Alexis were engineering majors at the University of Oslo, taking the physics course as part of their electronics, informatics and technology bachelor's degree. They based their project, which investigated lightning safety in cars, on one of the “seed” programs provided to UiO students. The program in question computationally simulated a lightning strike on flat ground, using certain electricity and magnetism concepts taught in the course, and one of the suggested investigation questions suggested that students use the simulation to investigate whether a car was actually a safe place to shelter during a lightning strike.

In their interview, the pair reported that they had been quite curious about this topic, and so had spent extensive time reading and writing about it. Their essay clearly reflected this interest, featuring an extensive and detailed theoretical introduction including several eye-catching graphics illustrating different aspects of lightning safety in cars. However, when the pair reached the step of computationally modeling how lightning would strike a car, they struggled to find a way to implement a car-like structure within the simulation they were building on. Additionally, assignment guidelines stated that students could expect to spend approximately ten hours on their project, and the pair realized that they had already reached that quota in the course of their background reading and writing. So, with the project deadline approaching, they left the provided code mostly untouched, pasted in a graphic from the textbook showing how they expected the simulation should look, and wrote a short summary and conclusion.

The pair described this decision as follows (translated from Norwegian):

**Amy:** *We wanted to actually try to program it so that it would show how lightning would strike our car. How lightning would strike it and then go around and down instead of through, but we don't know if we had any, yeah…*

**Alexis:** *It's only 10 hours you're supposed to use, we used more, but yeah…*

**Amy:** *Yeah, and we had a lot of other things to do, so then it was like, okay, we can rather… And we found this picture in the textbook, so it was really this that we were trying to simulate, but I don't know. Maybe if we had had a little more time, or… then we could have tried to get a bit further.*

In the end, the pair fulfilled the necessary criteria to successfully pass the project. However, because they used one of the default investigation questions instead of defining their own and made minimal modifications to their code, their essay demonstrated low levels of agency (scoring a 0 or 1 out of 2) across several major categories (*Programming and Data Processing*, *Investigation*, and ⅔ of the sub-categories of *Modeling and Data Analysis*). The students' description of their difficulties with (and ultimate abandonment of) the computational modeling parts of the project align with this characterization. At the same time, the students clearly spent a great deal of time and energy researching and writing about their chosen topic, and their scores in *Writing Genre and Polish*, *Graphics, Images and Illustrations*, and *External Sources* reflected this. This case, then, serves as an example of students who had an overall low level of agency, according to our codebook, due to the fact that they focused their attention exclusively on the communicative aspects of their computational essay.

**Medium agency: Iggy defines a novel question but performs a surface-level analysis**

Iggy was a 2nd-year computer science student at Michigan State University with an interest in data science. When the time came to choose a project, he initially struggled to settle on a topic, but eventually decided on an analysis relating to his interest in baseball:

> **Iggy:** *When we got to the project brainstorming phase, I'm not a very creative person, it was something that, it took a lot of days just sitting there and being, "No, no. That's not creative." But my biggest interest is baseball. My goal down the line for a career... [...] I'd like to do Sabermetrics, data analysis, looking at different stats, building algorithms, looking for patterns and such in a baseball context. Looking up how do we get the best players, or the best team based on a data-driven look than a scouting look. So I wanted to do something like that for my project.*

Based on this interest, he chose to analyze several of the most commonly-used baseball statistics to see which was most correlated with a team's winning percentage. In his essay, after downloading and loading in several teams' datasets from baseball-reference.com, he calculated and plotted the correlation coefficient between different teams' win percentages and 10 different variables such as batting average, offensive strikeouts, and offensive walks. After summarizing these statistics, he concluded his essay with a summary of the various correlations and a brief discussion of the difference between correlation and causation.

Iggy's essay received exclusively 1s on all categories, with the exception of investigation question and external sources (where it received 2s). Thus, this essay provided an example of an assignment in which the student had fulfilled all of the criteria, but only demonstrated agency in two areas: the question he chose to investigate, and the sources used to investigate it (including his chosen dataset). When interviewed, Iggy explicitly reflected on this point, describing how the most difficult parts of the project had been finding an appropriate question and dataset:

> **Interviewer:** *Okay. So you mentioned that the hardest part was choosing a topic. What made that so hard?*
>
> **Iggy:** *Kind of like I just said, aside from the fact that I'm really not a creative mind, being able to find data that we can use for our topic is one of the hardest things because we have an open internet with lots of different information and data on it, but to find exactly what we're looking for, something that we can modify to be what we're looking for can be a challenge, especially for what I was doing which was looking at inning by inning scores of baseball games.*

Once he had identified his question and dataset, Iggy deliberately chose a simple, surface-level analysis for the remainder of his project:

> **Interviewer:** *So what motivated you to go this route of just looking at, I think you said, ten different statistics, and looking at correlation and then making that the main focus of the project, rather than potentially other... How did you decide on that route?*

> **Iggy:** *To be honest, it was the simplest way to do it. I didn't want to make anything harder than it had to be. So with this project, which was basically just trying to find which statistic is the strongest, there wasn't really any methods we learned in class I could think of, going through everything, that measures strictly correlation.*

Iggy's reference to not being a "creative mind", along with his explicit reference to looking for "the simplest way to do it" support our finding that this case represents a medium-level of epistemic agency.

**High agency: Margaret and Edward build a detailed model of railgun dynamics based on real-life examples**

Margaret and Edward were 2nd-year students at the University of Oslo, majoring in materials science and physics respectively. For their project, the pair decided to build off of one of the provided simulations, which analyzed the motion of a projectile being launched out of a railgun. However, rather than using one of the suggested investigation questions, they instead chose to explore the effects of various phenomena they had learned about in a mechanics course they had taken the previous semester. As Edward explained during their interview,

> **Edward:** *We didn't really know how to formulate a good question in the start, but we knew we wanted to use the railgun as a model. Without being too unrealistic we wanted a realistic use of the railgun. We wanted to use the things we learned in mechanics of*

*PhysMech [introductory mechanics] to model how it would move and forces would act upon the projectile. That's basically what we tried to do.*

To pursue this question, the pair began by researching actual railguns to try to find physically meaningful parameters for their simulation. They summarized their findings in the introduction of their essay, embedding a dramatic video clip of an actual railgun firing for illustration. They then built up a simple model of a railgun using well-documented and well-explained code, much of which was compartmentalized in functions. During their interview, they explicitly referenced this emphasis on good coding practices:

> **Margaret:** *We wanted coding that was efficient, even though sometimes I felt, at least, that we weren't that good in programming and we could write a program that would be even more efficient, but our focus was on making the program as efficient as possible.*

Once they had created a simple working model, they refined it in several ways: implementing a function to allow them to find the appropriate angle necessary to launch the projectile a specified horizontal distance, implementing air resistance into their model, and exploring the effects of the coriolis force on projectile trajectory. They illustrated their results with multiple plots, and explicitly discussed their efforts to provide useful illustrations for readers:

> **Edward:** *We wanted to present the visualization, the different distances it reached, because you could see the effect of the air resistance, which is quite big when the speed is such a high number, right?*

In their conclusion, they discussed the assumptions they made along the way, the limitations these assumptions put on the analysis, and provided an extensive list of works cited from a variety of sources.

> **Edward:** *Then we talked a little about results and then about the real life implications that might actually happen.* [...] *Of course there's a lot of problems because of the forces on the…* [rails] *yeah, and also the temperature as well, which develops in such a fast cannon, of course. Also a lot of upsides because you can shoot it as many times as you want. You don't need heavy ammunition or... there's no chemistry involved, no explosions, right? It would be a lot easier if they get it to work. And then sources, of course, at the end. They didn't really say they want sources, so we just put the links to the videos and we didn't really talk about them, but it's there.*

This case exemplifies high levels of epistemic agency across most of the dimensions captured by our analysis. The students defined a novel question, based in a real-life context, and built a sophisticated multi-step model to investigate it. Their code was well explained, structured, and documented. Their essay explicitly attended to readability and polish, including multiple types of visualizations (video and plots) to illustrate phenomena and results. As a result, their essay received scores of 2 on all categories, save for code efficiency and elegance (since several parts of the model used copy-and-pasted code) and new packages and coding tools (since they only used tools that had been taught in the course).

These three narratives serve to illustrate the breadth of ways students chose to approach the computational essay project. Thus, they begin to address our first research question, as well as providing evidence that our codebook does, in fact, capture meaningful distinctions between student approaches. To further address both our first and second research questions, we now turn to a broader analysis of our entire coded dataset.

## Results Part 2: Overall trends in the computational essay dataset

In this section we unpack the large-scale trends revealed by this analysis and describe potential connections to essay characteristics and educational design decisions across the two featured institutions.

### Aggregate trends

When we combined all essays from both UiO and MSU into one group, our analysis revealed several interesting large-scale trends and patterns in the way students demonstrated epistemic agency. These trends are visible in Figure 3, which shows the percentage of students in the entire dataset who received scores of 0, 1, and 2 in each category.

As can be seen in Figure 3, in most categories/sub-categories the majority of students demonstrated a level of performance that would be considered sufficient for a passing grade (1 or 2). This result can also be seen in Figure 2, which shows that only a few essays demonstrated low levels of epistemic agency across multiple categories. This result, however, is not necessarily surprising as both institutions provided students with substantial scaffolding and support, including rubrics that specified the criteria necessary for a passing grade.[1] The only category with a significant number of 0 scores was "External Sources" (MD3); however, this was likely due to the fact that students at the University of Oslo were encouraged but not explicitly required to cite external sources. This effect can also be seen in Figure 4.

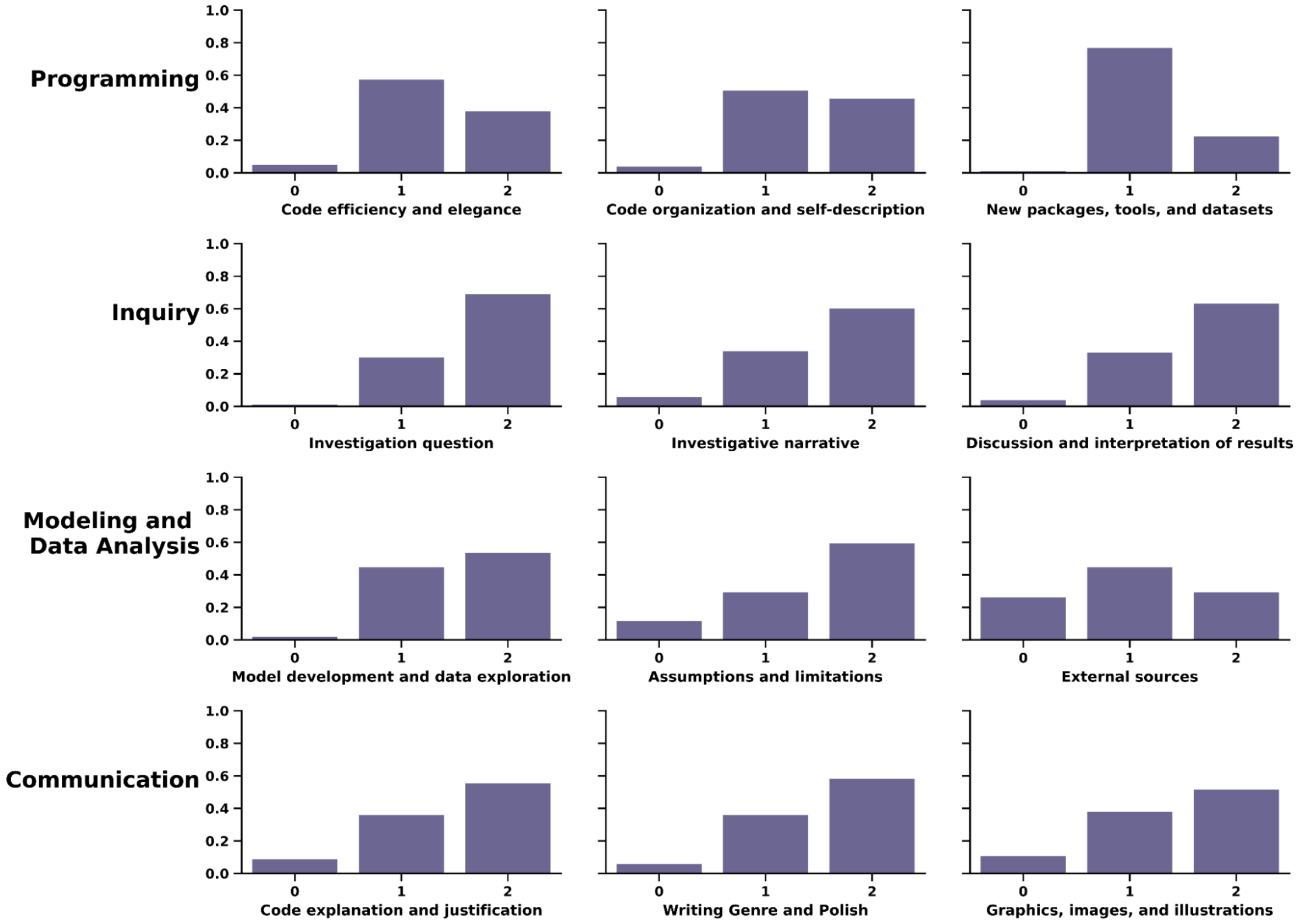

**Figure 3:** Percentages of students receiving specific epistemic-agency scores across each of the 12 sub-categories, aggregating data from the University of Oslo and Michigan State University.

[1] We note, however, that this analysis only includes data from consenting students, which could also create a self-selection bias (i.e., students who failed the project may have been less likely to consent to data collection).

Figure 3 also shows that for certain sub-categories (e.g., all of the *Inquiry* sub-categories) high levels of student agency was often the default rather than the exception. A qualitative review of the collected computational essays supports this conclusion. For instance, many of the essay topics were personally relevant to the students. This fact was often made clear by explicit references in the introduction (e.g., an essay on baseball statistics which began "I have always enjoyed playing and watching baseball. It is a data-driven sport that…") or the topic of the essay (e.g., "Could we use the concept of a rail gun to make a Space Elevator?"). Some essays also displayed clear markers of student agency throughout the narratives: for example, many essays from the Norwegian dataset included an analysis of the approximate cost of electricity for simulated tasks, like the aforementioned railgun-driven space elevator. Because Norway's economy is heavily energy-driven, this emphasis likely represents an example of students' home culture being integrated into the computational essay task.

We also see a tentative connection between several of these trends and the scaffolding and expectations provided for the students. For example, students at MSU were explicitly required to define their own investigation question (including finding their own dataset if they were doing a data-driven project) and could not complete their project without doing so. For these students, the only reason that they would receive a "1" in the corresponding sub-category (I1) would be if they either failed to include their question in their essay or had a particularly ill-defined question. We explore these connections in greater detail in the next section.

Returning to our first research question, these results show that computational essays were a successful vehicle for the redistribution of epistemic agency to students at both institutions included in the study. As can be seen from Figure 2, nearly every student in the dataset had at least one area in which they demonstrated high levels of agency, and many demonstrated high agency across multiple categories. Moreover, as can be seen from Figures 2 and 3, every sub-category showed high levels of agency in at least 20% of the dataset. However, differences in course context, expectations, and scaffolding clearly also had an effect on student performance. To unpack these differences, we must disaggregate the dataset by institution.

**Differences between the two institutions**

By splitting the dataset by institution, we were able to make more direct comparisons between the degrees of agency demonstrated by the students across these two educational contexts. A normalized histogram of the codes, split by institution, is shown in Figure 4.

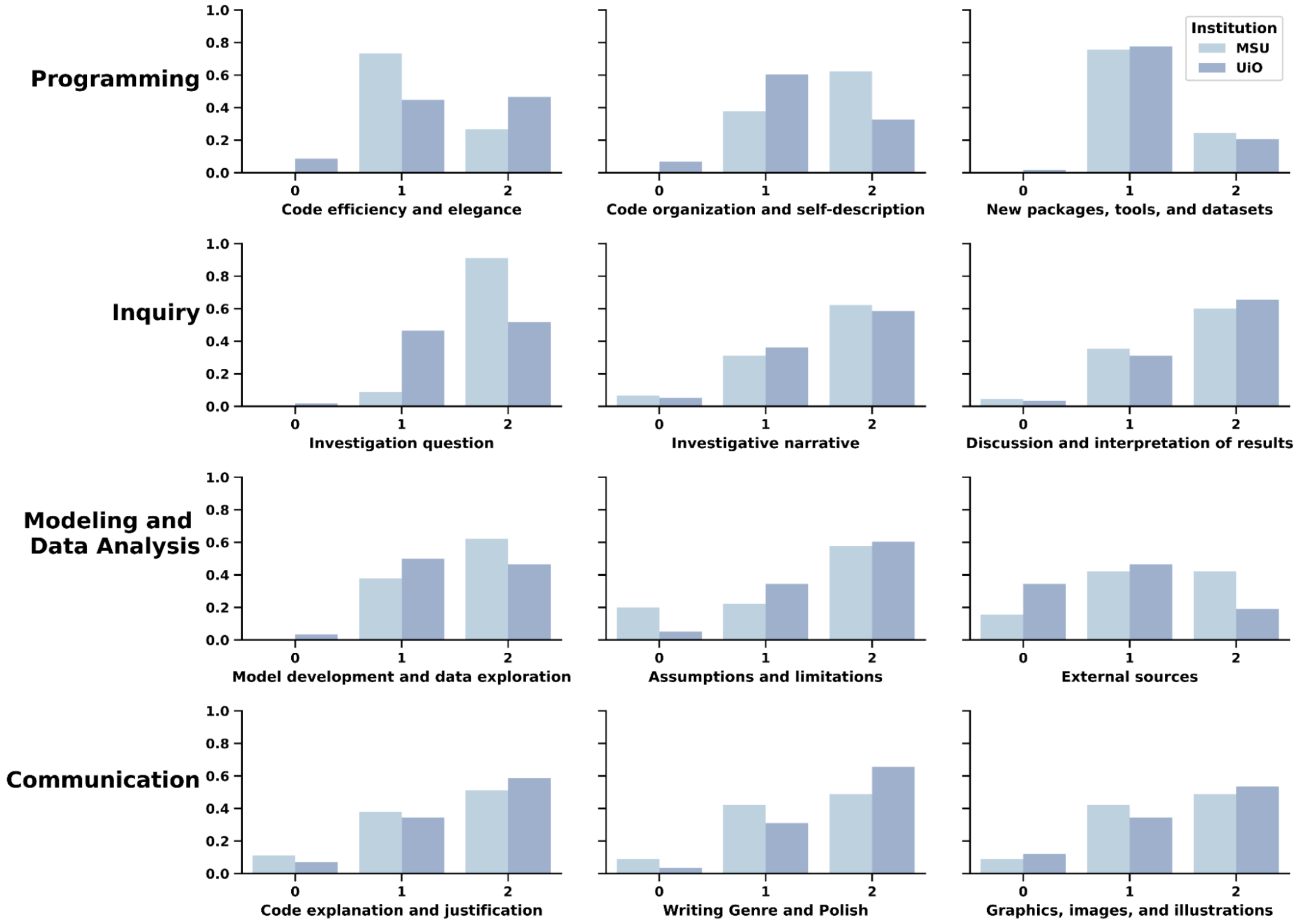

**Figure 4:** Percentages of students receiving specific epistemic-agency scores across each of the 12 sub-categories, separated by institutions (Michigan State vs. University of Oslo).

Looking at this comparison, we can make several observations. Several sub-categories appear to be fairly isomorphic across the two institutions (e.g., P3, I2, I3, C1, and C3). We suggest that these similarities indicate points of alignment between the two educational designs: for example, both encouraged students to spend time presenting and justifying their analysis using both a mix of text and code (C1) and graphics, images, and illustrations (C3). Similarly, both designs encouraged students to present their results in a narrative form (I2), including discussion of relevant project results (I3). However, neither design explicitly required students to learn new programming tools for the project (P3) although some students did, in fact, do so.

Other categories show clear differences across the two institutions: these include P1, P2, I1, MD3, and C2. We interpret these differences as reflecting differences between the two educational designs. For example, UiO students demonstrated higher overall levels of agency on *Code Efficiency and Elegance* (P1) and *Writing Genre and Polish* (C2). We suspect that the higher degree of agency in code efficiency (P1) is likely due to the fact that most of the UiO students had taken an object-oriented programming course during their first semester, giving them a foundation in efficient computing prior to the course in question. In contrast, most of the MSU students were learning to code for the first time and consequently have likely not yet developed the same levels of programming skill. Similarly, the UiO students' emphasis on writing genre and polish (C2) may stem from the fact that UiO students were presented with a gallery of highly-polished computational essays when they were first introduced to the assignment which were meant to raise their expectations and encourage them to produce similarly polished assignments. MSU students, in contrast, are shown a more limited number of example essays, of varying levels of quality.

The three sub-categories where MSU shows greater student agency (P2, I1, and MD3) also seem to result from differences in the two educational designs. As previously discussed, all MSU students are required to define their own investigation question, which likely explains the greater emphasis on the *Investigation question* sub-category (I1). UiO students, in contrast, were given the option to work off of a set of predefined questions, and many students chose this option (resulting in a lower score in this category). We suspect the greater emphasis by MSU students on *Code organization and self-description* (P2) is due to the fact that MSU students work within the Jupyter Notebook framework from the first day of class and are presented with numerous assignments that model best-practices for notebook use and code documentation. UiO students, on the other hand, have less training in documenting code, and many UiO students reported this project as their first exposure to Jupyter notebooks. Finally, MSU students are explicitly required to include external sources in their essays (MD3) including citing any external dataset included in their work. At UiO, the use of external sources was encouraged but not required to pass.

These results help address our second research question: that is, they show how differences in implementation and scaffolding can help to influence (or direct) students towards certain categories of agency. From one perspective, this is unsurprising: if one provides students with the option to build off of a suggested investigation question, there will naturally be fewer students who define their own, more novel questions. Similarly, if students have a strong background in code documentation or object-oriented programming, it seems reasonable to expect that this will be reflected in the choices they make (and the way they demonstrate agency) in their projects. However, these results imply that educators must be cognizant of these factors when implementing open-ended, project-based assignments like computational essays. If, for example, they wish to assess students on certain aspects of programming, inquiry, modeling, or communication, it is important to take into account student background, comfort, scaffolding, and grading criteria.

**Association analysis to determine overlap between theoretical categories**

In the final step of our analysis, we were interested to find out whether the theoretical distinctions we had made between different sub-categories of epistemic agency were robust in practice. In other words, certain sub-categories (such as I3: *Discussion and interpretation of results* and MD2: *Assumptions and limitations*) were theoretically distinct, but often seemed to overlap in practice (i.e., a discussion paragraph where students both summarize results and describe the limitations of their analysis). This begs the question: are these in fact meaningful distinctions, or are these categories practically indistinguishable?

To answer this question, we performed an association analysis by computing the bias-corrected Cramér's V (Bergsma, 2013; Cramer, 1999; Kearney, 2017) across all categories, splitting the dataset by institution. We assumed that, if certain categories were in fact indistinguishable, they would appear as strong associations that were similar across both institutions. If, on the other hand, associations appeared at one institution but not the other, it more likely indicated that the specifics of implementation, scaffolding, or expectations at that institution led to the overlap (and consequently that the theoretical distinction was, in fact, meaningful). The resulting association plot is shown in Figure 5.

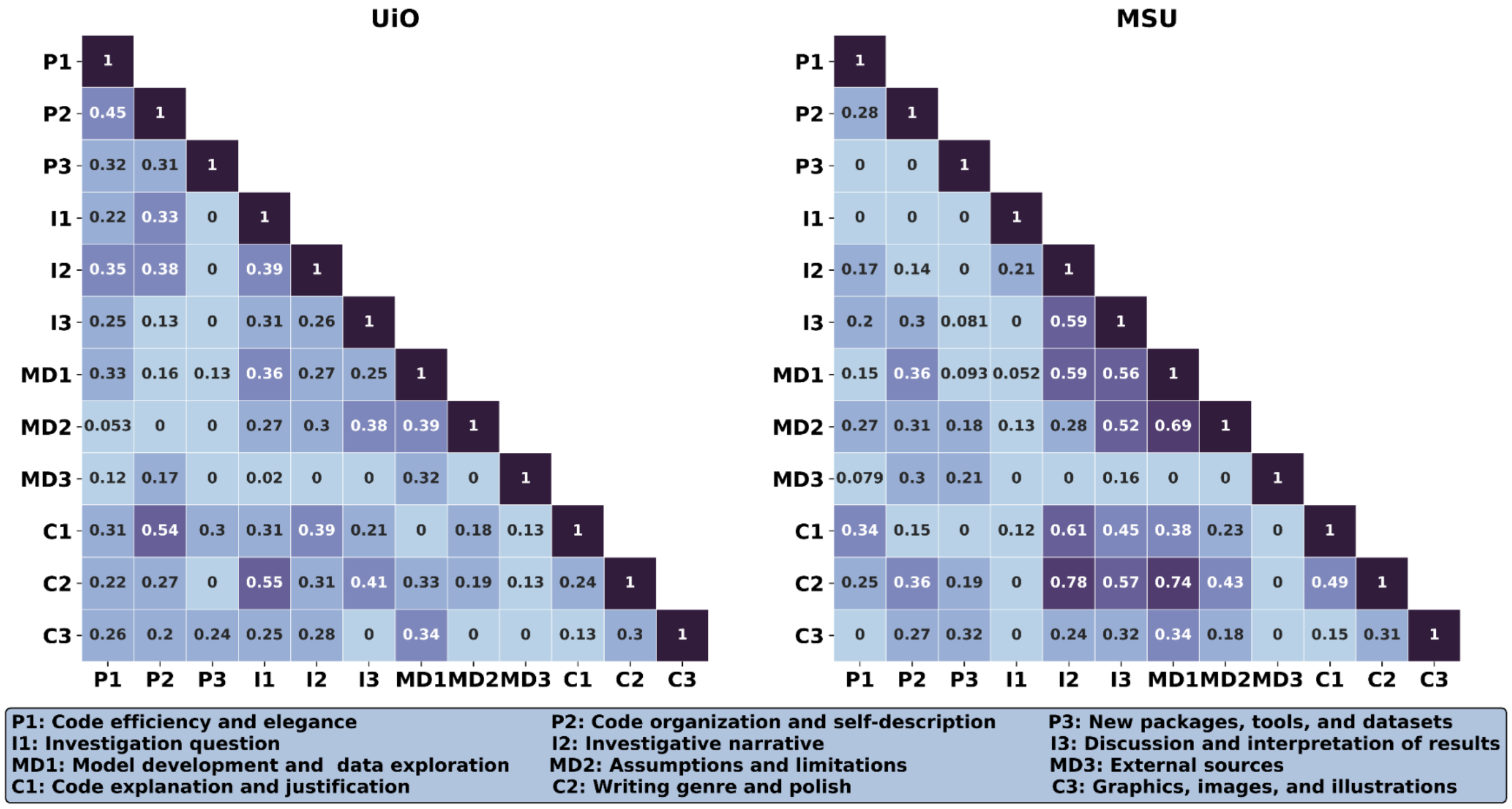

**Figure 5:** Heatmap of category association based on the bias-corrected Cramér's V, split by institution.

Using the threshold proposed by Cohen (1988) of V > 0.5 to establish a strong association, this analysis suggests that the theoretical distinctions we have made between different sub-categories of epistemic agency are, in fact, meaningful. At UiO, there are only two such strong associations, between C1/P2, and C2/I1. MSU shows significantly more strong associations, such as an association between C2/I2, C2/MD1, and MD1/MD2; however these do not overlap with the associations from UiO. Thus, even though certain categories may overlap in certain implementations or types of essays, it appears that the theoretical distinctions themselves are justified.

Moreover, we can hypothesize specific differences in student background or computational essay implementation that might account for these associations. For example, the association between *Writing genre and polish* (C2) and *Investigation question* (I1) at UiO may be related to the fact that UiO students were given the option to build their projects off the provided "seed" programs, including being able to use the suggested investigation questions provided in those notebooks (e.g., Amy and Alexis). In many cases, students who chose to use one of these programs and associated questions seemed to spend less effort on the writing in their essays, perhaps assuming that their audience (the graders and fellow students) were already familiar with the concepts underlying their projects. At MSU the association between *Discussion and interpretation of results* (I3) and *Investigative narrative* (I2) may be related to the fact that many of the MSU essays focused on statistical analysis of publicly-available datasets; in these analyses, the discussion and interpretation of results was often woven into the investigative narrative rather than being a detailed stand-alone section (as was common in model-based UiO essays). The MSU association between *Assumptions and limitations* (MD2) and *Model development and data exploration* (MD1) seems to be related to a trend wherein students who chose to pursue data-driven projects often carried out somewhat superficial analyses which focused on quickly visualizing data and capturing surface-level trends or associations without considering the associated limitations. Iggy's narrative exemplifies this trend.

## Discussion and implications

In summary, and returning to our first research question, we argue that computational essays provide significant affordances for helping redistribute epistemic agency to students. First, they allow students to define novel, personally-relevant questions, answer them with scientifically-authentic tools, and communicate the analysis in an authentic way. These questions can be general (as with Amy and Alexis) or specific (as with Margaret and Edward); tightly connected with a scientific discipline (as with both of the UiO groups) or based entirely on one's own interests (as with Iggy). Put differently, computational essays allow students to adapt and leverage the tools of scientific inquiry for their own purposes, potentially independent of teacher oversight. Second, computational essays are flexible, allowing students the freedom to choose which aspects of the process—coding, inquiry, modeling and data analysis, or presentation—they wish to focus on. Different students in our study naturally gravitated toward different aspects of computational essays, as can be seen from Figure 2; however, all of these aspects are authentic to the discipline of science, meaning that computational essays allow for multiple productive modes of engagement with authentic scientific practices. Third, they provide students with a significant degree of control over the inquiry process: students can choose how far they wish to take their analysis and when they consider their question to be answered. Some students might be comfortable with surface-level analysis (Iggy) whereas others may wish to build detailed models incorporating multiple effects (Margaret and Edward).

Addressing our second research question, our analysis also shows that factors like student background, assignment expectation, and scaffolding can have a significant impact on the ways in which students take up epistemic agency when writing computational essays. In terms of student background, students at UiO had experience with object-oriented programming which likely led to the greater emphasis on code efficiency; at MSU, students had a stronger background in code documentation and notebook use, leading to a greater emphasis on code organization and description. In terms of assignment expectations, students at MSU were explicitly expected to define novel research questions and consult (cite) external sources, leading to significantly higher levels of demonstrated agency across these categories. As for scaffolding, the fact that students at UiO were provided with a gallery of polished example essays likely contributed to the greater degree of polish in their written reports. These differences, however, again demonstrate the flexibility of computational essays as a medium: by adjusting the parameters of the educational design or adapting the assignment to student backgrounds, students can be encouraged to take up agency in a variety of different ways.

There are several impactful results from this study. First, this study contributes to the growing theory of epistemic agency in science education. That is, in the course of our analysis, we have taken the broad theoretical construct of epistemic agency and operationalized it to analyze the specific affordances of computational essays in science education. We posit that this codebook, and the conceptual framework embedded in it, may be used and adapted to evaluate the degree of agency demonstrated within other kinds of open-ended computational science projects as well. That is, the theoretical categories of programming, inquiry, modeling and data analysis, and communication are certainly applicable to many different types of computational projects, whether they are written in a computational essay format or not. Furthermore, we have provided evidence that the theoretical distinctions made in this codebook are, in fact, meaningful when applied across different implementations of computational essays.

Second, we have demonstrated that computational essays can be a useful tool in redistributing epistemic agency to students, granting them cognitive authority over their learning. As we have argued, postsecondary science education is in significant need of innovations to help redistribute epistemic agency to students—computational essays, and perhaps the adoption of scientific computing more generally, represent just such an innovation. As we have seen, computational essays allow modes of coursework that go beyond the reproduction of established scientific knowledge; in fact, they allow students to make analyses, observations, and discoveries that may well be completely new. And, because computational essays are an authentic and frequently-used mode of communication by professional scientists, we predict that their use will only increase in upcoming years, especially as computation continues to be integrated into various scientific disciplines.

At the same time, our results speak to the flexibility of computational essays, which allow students to choose whether they wish to focus on programming, inquiry, modeling and data analysis, communication, or some combination of the above. This flexibility certainly presents opportunities, but also requires educators to be cognizant that different students may require varying levels of support to successfully meet all expectations embedded in such an assignment. As seen with the example of Amy and Alexis, if students do not receive the requisite support (or oversight) they may neglect key parts of the assignment and thereby miss key learning goals.

There are, of course, some clear limitations to this study. First, there is the fact that we have presented data from only two courses from two institutions from a single semester. A greater variety of data would certainly help us to strengthen and refine the claims we have made here, especially regarding the connections between educational context and epistemic agency demonstrated by students. Second, many of our conclusions about student agency are based on our own judgments as practicing scientists and data analysts. They thus represent a top-down view of the phenomenon under study—a phenomenon that is, at its core, individual to learners. This top-down view is to some degree coherent with our theoretical framework, since one of the key aspects of epistemic agency in science is authenticity to the discipline. However, we must be cognizant that students come from a variety of backgrounds, and what may appear to us as, for example, a mediocre project may in fact have been very challenging (and required a great deal of agency) for the student who did it.

Third, much of our analysis comes from evaluation of student artifacts, which naturally has the limitation of only showing a finished product. We thus do not have any insight into the students' thought processes as they pursued the majority of these projects. We tried to offset this limitation, to some degree, by interviewing subsets of students to ask about their thought processes. However, even these interviews are retrospective and do not reflect the rich messiness of the inquiry process.

Finally, we must acknowledge the effects of self-selection bias in our dataset. We were only able to collect artifacts from a subset of students taking each course. It is possible, even likely, that the students who consented to let us use their work were those who were most secure or proud of the work they had done. Thus, we suspect our results might overestimate the spectrum of agency present in the courses as a whole.

Despite these limitations, we find these results encouraging. Broadly speaking, the results from this project suggest that computation can be an important tool in the ongoing effort to make science learning more authentic to the scientific disciplines and grant students more agency in their learning.

## Conclusion

It has long been known that the structures within schools set the expectations for student and teacher roles, which determines what kinds of knowledge are valuable, who holds that knowledge, and how it is constructed (Stroupe, 2014; Warren & Rosebery, 1995). In this regard, we argue that computational essays, when used for open-ended, inquiry-based, student-driven projects, may act as a structure for the redistribution of epistemic agency to science students. Uptake of this structure will certainly require adjustments to the current paradigm of teaching in higher education STEM, such as a decreased focus on students achieving one singular outcome or right answer. However, given the potential benefits of allowing students to experience authentic scientific inquiry, using authentic scientific methods, and learning an authentic mode of scientific communication, these seem like relatively minor tradeoffs—especially if computational essays are used as a single assignment within a larger course.

Although we have presented two examples of computational essay implementation, we are greatly interested to see more. Computational essays have been discussed, at least theoretically, for the last two decades (diSessa, 2000; Somers, 2018; Wolfram, 2017). Now, however, we are finally seeing what it looks like when this genre begins to make its way into educational contexts. We see computational essays as a solid foundation for bringing in more principles of ambitious science teaching into the higher education sphere, and are excited for the possibility that computational essays may act as a site for this type of change.

## Acknowledgments

We would like to thank Sebastian Gregorius Winther-Larsen, Marcos (Danny) Caballero, David Stroupe, and Crina Damşa for their support and feedback in this project. This study was funded by Norges Forskningsråd Project No. 288125 “International Partnership for Computing in Science Education” and by the Norwegian Agency for International Cooperation and Quality Enhancement in Higher Education (DIKU) which supports the Center for Computing in Science Education.

## References

American Association For The Advancement of Science. (2011). *Vision and change in Undergraduate Biology Education: A call to action, Final Report*. *3*, 2018.

Association of American Universities. (2017). *Progress Toward Achieving Systemic Change: A Five-Year Status Report on the AAU Undergraduate STEM Education Initiative*. https://www.aau.edu/sites/default/files/AAU-Files/STEM-Education-Initiative/STEM-Status-Report.pdf

Barton, A. C., & Tan, E. (2010). We be burnin’! agency, identity, and science learning. *Journal of the Learning Sciences*, *19*(2), 187–229.

Bergsma, W. (2013). A bias-correction for Cramér’s V and Tschuprow’s T. *Journal of the Korean Statistical Society*, *42*(3), 323–328.

Berland, L. K. (2011). Explaining Variation in How Classroom Communities Adapt the Practice of Scientific Argumentation. *Journal of the Learning Sciences*, *20*(4), 625–664.

Berland, L. K., & Crucet, K. (2015). Epistemological Tradeoffs: accounting for context when evaluating epistemological sophistication of student engagement in scientific practices. *Science Education*.

Berland, L. K., & Reiser, B. J. (2011). Classroom communities’ adaptations of the practice of scientific argumentation. *Science Education*, *95*(2), 191–216.

Berland, L. K., Schwarz, C. V., Krist, C., Kenyon, L., Lo, A. S., & Reiser, B. J. (2015). Epistemologies in practice: Making scientific practices meaningful for students. *Journal of Research in Science Teaching*. https://doi.org/10.1002/tea.21257

Blakeslee, A. M. (1997). Activity, Context, Interaction, and Authority: Learning to Write Scientific Papers in Situ. *Journal of Business and Technical Communication*, *11*(2), 125–169.

Brownell, S. E., Kloser, M. J., Fukami, T., & Shavelson, R. (2012). Undergraduate Biology Lab Courses: Comparing the Impact of Traditionally Based “Cookbook” and Authentic Research-Based Courses on Student Lab Experiences. *Journal of College Science Teaching*, *41*(4), 36–45.

Caballero, M. D., & Merner, L. (2018). Prevalence and nature of computational instruction in undergraduate physics programs across the United States. *Physical Review Physics Education Research*, *14*, 020129–020129.

Center for Computing in Science Education. (2019). *Computational Essays from the University of Oslo*. https://uio-ccse.github.io/computational-essay-showroom/index

Chinn, C. a., & Malhotra, B. a. (2002). Epistemologically Authentic Inquiry in Schools: A Theoretical Framework for Evaluating Inquiry Tasks. *Science Education*, *86*(2), 175–218.

Cohen, J. (1988). *Statistical Power Analysis for the Behavioral Sciences (2nd ed.)*. Lawrence Erlbaum Associates.

Cooper, M. (1994). Cooperative Chemistry Laboratories. *Journal of Chemical Education*, *71*(4), 307.

Cooper, M., & Klymkowsky, M. (2013). Chemistry, Life, the Universe, and Everything: A New Approach to General Chemistry, and a Model for Curriculum Reform. *Journal of Chemical Education*, *90*(9), 1116–1122.

Cramer, H. (1999). *Mathematical methods of statistics (PMS-9)*. Princeton University Press. https://doi.org/10.1515/9781400883868

Damşa, C. I., Kirschner, P. a., Andriessen, J. E. B., Erkens, G., & Sins, P. H. M. (2010). *Shared Epistemic*

*Agency: An Empirical Study of an Emergent Construct* (Vol. 19).

Denning, P. J., & Tedre, M. (2019). *Computational Thinking*. MIT Press.

diSessa, A. A. (2000). *Changing minds: Computers, learning, and literacy*. Mit Press.

Dounas-frazer, D. R., Bogart, K. L. V. D., Stetzer, M. R., & Lewandowski, H. J. (2016). Investigating the role of model-based reasoning while troubleshooting an electric circuit. *Physical Review Physics Education Research*, *12*, 1–20.

Dounas-Frazer, D. R., & Lewandowski, H. J. (2016). Nothing works the first time: An expert experimental physics epistemology. *Physics Education Research Conference*, *June*, 4–4.

Dounas-Frazer, D. R., & Lewandowski, H. J. (2018). The Modelling Framework for Experimental Physics: Description, development, and applications. *European Journal of Physics*, *39*(6). https://doi.org/10.1088/1361-6404/aae3ce

Feinstein, A. R., & Cicchetti, D. V. (1990). High agreement but low kappa: I. The problems of two paradoxes. *Journal of Clinical Epidemiology*, *43*(6), 543–549.

Gwet, K. L. (2014). *Handbook of Inter-Rater Reliability, 4th Edition: The Definitive Guide to Measuring The Extent of Agreement Among Raters*. Advanced Analytics, LLC.

Henderson, C., Beach, A., & Finkelstein, N. (2011). Facilitating change in undergraduate STEM instructional practices: An analytic review of the literature. *Journal of Research in Science Teaching*, *48*(8), 952–984.

Henderson, C., Dancy, M., & Niewiadomska-Bugaj, M. (2012). Use of research-based instructional strategies in introductory physics: Where do faculty leave the innovation-decision process? *Physical Review Special Topics - Physics Education Research*, *8*(2), 020104–020104.

Holmes, N. G. (2020). *Why Traditional Labs Fail… and What We Can Do About It* (J. J. Mintzes & E. M. Walter (eds.); pp. 271–290).

Holmes, N. G., Keep, B., & Wieman, C. E. (2020). Developing scientific decision making by structuring and supporting student agency. *Physical Review Physics Education Research*, *16*(1), 10109.

Holmes, N. G., & Wieman, C. E. (2016). Examining and contrasting the cognitive activities engaged in undergraduate research experiences and lab courses. *Physical Review Physics Education Research*, *12*(2), 020103–020103.

Holmes, N. G., & Wieman, C. E. (2018). Introductory physics labs: We can do better. *Physics Today*, *71*(1), 38–45.

Irving, P. W., Mcpadden, D., & Caballero, M. D. (2020). Communities of practice as a curriculum design theory in an introductory physics class for engineers. *Physical Review Physics Education Research*, *16*(2), 20143.

Irving, P. W., Obsniuk, M. J., & Caballero, M. D. (2017). P3: a practice focused learning Environment. *European Journal of Physics*, *38*, 1–21.

Kearney, M. W. (2017). Cramér's V. In M. Allen (Ed.), *The SAGE Encyclopedia of Communication Research Methods*.

Kloser, M. J., Brownell, S. E., Chiariello, N. R., & Fukami, T. (2011). Integrating teaching and research in undergraduate biology laboratory education. *PLoS Biology*, *9*(11), e1001174.

Kluyver, T., Ragan-Kelley, B., Pérez, F., Granger, B., Bussonnier, M., Frederic, J., Kelley, K., Hamrick, J., Grout, J., Corlay, S., Ivanov, P., Avila, D., Abdalla, S., Willing, C., & Jupyter Development Team. (2016). *Jupyter Notebooks—a publishing format for reproducible computational workflows* (F. Loizides & B. Schmidt (eds.); Vol. 16, pp. 87–90). IOS Press.

Ko, M. L. M., & Krist, C. (2019). Opening up curricula to redistribute epistemic agency: A framework for supporting science teaching. *Science Education*, *103*(4), 979–1010.

Lehrer, R., & Schauble, L. (2007). Scientific thinking and science literacy. In *Handbook of Child Psychology*. John Wiley & Sons, Inc. https://doi.org/10.1002/9780470147658.chpsy0405

Livingstone, D. N. (2010). *Putting Science in Its Place: Geographies of Scientific Knowledge*. University of Chicago Press.

Magana, A. J., Falk, M. L., Vieira, C., & Reese, M. J. (2016). A case study of undergraduate engineering students' computational literacy and self-beliefs about computing in the context of authentic practices. *Computers in Human Behavior*, *61*, 427–442.

Miller, E., Manz, E., Russ, R., Stroupe, D., & Berland, L. (2018). Addressing the epistemic elephant in the room: Epistemic agency and the next generation science standards. *Journal of Research in Science Teaching*, *55*(7), 1053–1075.

Moskovitz, C., & Kellogg, D. (2005). Primary Science Communication in the First-Year Writing Course. *College Composition and Communication*, *57*(2), 307–334.

NGSS Lead States. (2013). *Next Generation Science Standards: For States, By States*. The National Academies Press.

Odden, T. O. B., & Burk, J. (2020). Computational Essays in the Physics Classroom. *Physics Teacher*, *58*(March), 252–255.

Odden, T. O. B., Lockwood, E., & Caballero, M. D. (2019). Physics Computational Literacy: An Exploratory Case Study Using Computational Essays. *Physical Review Physics Education Research*, *15*(2), 20152–20152.

Odden, T. O. B., & Malthe-Sørenssen, A. (2021). Using computational essays to scaffold professional physics practice. *European Journal of Physics*, *42*(015701), 1–22.

Olson, S., & Riordan, D. G. (2012). Engage to excel: Producing one million additional college graduates with degrees in science, technology, engineering, and mathematics. Report to the President. *Executive Office of the President*. http://files.eric.ed.gov/fulltext/ED541511.pdf

Rudolph, J. L. (2005). Epistemology for the Masses: The Origins of "The Scientific Method" in American Schools. *History of Education Quarterly*, *45*(3), 341–376.

Rule, A., Tabard, A., & Hollan, J. (2018). *Exploration and Explanation in Computational Notebooks*.

Sandi-Urena, S., Cooper, M., & Stevens, R. (2012). Effect of Cooperative Problem-Based Lab Instruction on Metacognition and Problem-Solving Skills. *Journal of Chemical Education*, *89*(6), 700–706.

Scardamalia, M. (2002). Collective Cognitive Responsibility for the Advancement of Knowledge. *Liberal Edcuation in a Knowledge Society*, 67–98.

Scardamalia, M., & Bereiter, C. (1991). Higher Levels of Agency for Children in Knowledge Building: A Challenge for the Design of New Knowledge Media. *The Journal of the Learning Sciences*, *1*(1), 37–68.

Silvia, D., O'shea, B., & Danielak, B. (2019). A learner-centered approach to teaching computational modeling, data analysis, and programming. *International Conference on Computational Science*, 374–388.

Somers, J. (2018). The Scientific Paper is Obsolete. Here's what's next. *Atlantic* , 1–19.

Stains, M., Harshman, J., Barker, M. K., Chasteen, S. V., Cole, R., DeChenne-Peters, S. E., Eagan, M. K., Jr, Esson, J. M., Knight, J. K., Laski, F. A., Levis-Fitzgerald, M., Lee, C. J., Lo, S. M., McDonnell, L. M., McKay, T. A., Michelotti, N., Musgrove, A., Palmer, M. S., Plank, K. M., … Young, A. M. (2018). Anatomy of STEM teaching in North American universities. *Science*, *359*(6383), 1468–1470.

Stroupe, D. (2014). Examining Classroom Science Practice Communities: How Teachers and Students Negotiate Epistemic Agency and Learn Science-as-Practice. *Science Education*, *98*(3), 487–516.

Stroupe, D., Caballero, M. D., & White, P. (2018). Fostering students' epistemic agency through the co-configuration of moth research. *Science Education*, *102*(6), 1176–1200.

Thompson, K. V., Chmielewski, J., Gaines, M. S., Hrycyna, C. A., & LaCourse, W. R. (2013). Competency-based reforms of the undergraduate biology curriculum: integrating the physical and biological sciences. *CBE Life Sciences Education*, *12*(2), 162–169.

Warren, B., & Rosebery, A. S. (1995). Equity in the future tense: Redefining relationships among teachers, students, and science in linguistic minority classrooms. *New Directions for Equity in*

*Mathematics Education*, 298–328.

Watkins, J., Coffey, J. E., Redish, E. F., & Cooke, T. J. (2012). Disciplinary authenticity: Enriching the reforms of introductory physics courses for life-science students. *Physical Review Special Topics - Physics Education Research*, *8*(1), 010112–010112.

Wieman, C. (2015). Comparative Cognitive Task Analyses of Experimental Science and Instructional Laboratory Courses. *Physics Teacher*, *349*(53). https://doi.org/10.1119/1.4928349

Windschitl, M., Thompson, J., Braaten, M., & Stroupe, D. (2012). Proposing a core set of instructional practices and tools for teachers of science. *Science Education*, *96*(5), 878–903.

Wolfram, S. (2017). *What Is a Computational Essay?* http://blog.stephenwolfram.com/2017/11/what-is-a-computational-essay/

Young, N. T., Allen, G., Aiken, J. M., Henderson, R., & Caballero, M. D. (2019). Identifying features predictive of faculty integrating computation into physics courses. *Physical Review Physics Education Research*, *15*, 010114–010114.